\documentclass{article}
\usepackage{epsfig,a4wide,amsfonts}
\newcommand{\be}{\begin{equation}}
\newcommand{\ee}{\end{equation}}

\newtheorem{theorem}{Theorem}[section]
\newtheorem{proposition}[theorem]{Proposition}
\newenvironment{proof}[1][Proof:]{\begin{trivlist}
\item[\hskip \labelsep {\bfseries #1}]}{\end{trivlist}}
\newcommand{\qed}{\nobreak \ifvmode \relax \else
      \ifdim\lastskip<1.5em \hskip-\lastskip
      \hskip1.5em plus0em minus0.5em \fi \nobreak
      \vrule height0.75em width0.5em depth0.25em\fi}

\begin{document}

\title{Designing arrays of Josephson junctions for specific static responses}
   \author{  J.G. Caputo$^{1,2}$ and L. Loukitch$^1$  \\
   {\normalsize \it 1) Laboratoire de Math\'ematiques, INSA de Rouen} \\
   {\normalsize \it B.P. 8, 76131 Mont-Saint-Aignan cedex, France } \\
   {\normalsize \it  2) Laboratoire de Physique th\'eorique et modelisation,} \\
   {\normalsize \it Universit\'e de Cergy-Pontoise and C.N.R.S., France}
   \date{\today}}

\maketitle

\begin{abstract}

We consider the inverse problem of designing an array
of superconducting Josephson junctions that has a given
maximum static current pattern as function of the applied magnetic
field. Such devices are used for magnetometry and as 
Terahertz oscillators. The model is a 2D semilinear elliptic operator
with Neuman boundary conditions so the direct problem is 
difficult to solve because of the multiplicity of solutions. 
For an array of small junctions in a passive region, 
the model can be reduced to a 1D linear partial differential equation
with Dirac distribution sine nonlinearities. For small junctions
and a symmetric device, the maximum current is 
the absolute value of a cosine Fourier series whose coefficients (resp. frequencies)
are proportional to the areas (resp. the positions) of the junctions. 
The inverse problem is solved by inverse cosine Fourier transform
after choosing the area of the central junction.
We show several examples using combinations of simple three
junction circuits. These new devices could then be tailored to meet
specific applications.

\end{abstract}

\section{Introduction}

The coupling of two Type I superconductors across a thin oxide layer is
described by the two Josephson equations \cite{josephson},
\begin{equation}\label{e1.1}
V=\Phi_0\frac{d\phi}{dt},~~~I = s J_c \sin(\phi).
\end{equation}
where $\phi$ is the phase difference between the two superconductors
in units of $\Phi_0=\hbar/2e$ the reduced flux quantum,
$V$ and $I$ are respectively, the voltage and current across
the layer, $s$ is the contact surface and $J_c$ is the critical current
density.
The Josephson equations and Maxwell's equations imply
the modulation of DC current by an external magnetic field in the static
regime (SQUIDs) and the conversion of AC current in microwave radiation
\cite{Barone,Likharev}. In all these systems there is a characteristic
length which reduces to the Josephson length $\lambda_J$, the 
ratio of the electromagnetic flux to the quantum flux $\Phi_0$ for standard 
junctions.
The behavior of a Josephson junction depends on its size compared to $\lambda_J$.
In small junctions the phase will not vary much except for large magnetic
fields. Long junctions on the contrary enable large variations of the phase
accommodating so-called "fluxons" or sine-Gordon 
kinks where the phase varies by $2\pi$
\cite{Barone}.

For many applications and in order to protect the junction, 
Josephson junctions are embedded in a so-called
microstrip line which is the capacitor made by the
overlap of the two superconducting layers. 
This is the "window geometry" where the phase
difference satisfies an inhomogeneous 2D damped 
driven sine Gordon equation \cite{cfv95} resulting
from Maxwell's equations and the Josephson 
constitutive relations (\ref{e1.1}).
For resonator
applications this design allows to couple the junctions in an array
to increase the
output power and adapt impedance for coupling the device to a transmission
line. In addition one can select some desirable dynamic features
like resonances \cite{cl05} and optimize the frequency response
over a given band for wave mixing applications \cite{Salez}.

Parallel arrays of Josephson junctions can be used in the
static regime as very fine magnetic field detectors. 
The maximal current $I_{max}$ which can cross the device 
(see Fig.~\ref{f1}) for a given magnetic field $H$, without any voltage 
($V=0$ the static regime) defines the $I_{max}(H)$ curve. 
The behavior of arrays of identical and equidistant small Josephson junctions
has been extensively studied \cite{Barone,Likharev}. The problem of
finding $I_{max}$ remains difficult to solve because of the multiplicity
of solutions due to the sine nonlinearity and the Neuman boundary conditions.

For fundamental reasons and applications it is interesting to work
with non-uniform arrays where the junction sizes and their spacings
can vary.
In \cite{cl05,cl06} we developed a continuous/discrete or long wave 
model where the phase variation is neglected in the junctions and
where the couplings between junction and surrounding microstrip are
correctly taken into account. In particular we consider the waves
between the junctions that are completely neglected by the classical
Resistive Shunted Junction (RSJ) lumped models \cite{Likharev}. 
Our approach allows to choose the distance between junctions and their 
area. In the same device we can model junctions with different 
areas and different current response, in particular 
$\pi$-junctions. This simple model allows to analyze in depth 
the statics of the device and this is not possible from
the 2D original equations \cite{cl06}. 
This long wave
approximation can be generalized to 2D to explain the behavior of
squids \cite{cg04}. 
In addition we obtain an excellent agreement with the complex
experimental $I_{max}(H)$ curves \cite{bcls06} using the very simple 
magnetic approximation introduced in \cite{cl06}.

For experimentalists, it is very useful to extract parameters
of the array from the $I_{max}(H)$ curve. 
For example it gives informations on the quality of the junctions.
Recent studies by Itzler and Tinkham examine how defects in the 
coupling affect this maximum current \cite{it95,it96}. This is 
important because high $T_c$ superconductors 
can be described as Josephson junctions where the
critical current density is a rapidly varying function of the position,
due to grain boundaries. Fehrenbacher et al\cite{fgb92} calculated
$I_{\rm max}(H)$ for such disordered long Josephson junctions and
for a periodic array of defects. The expressions obtained are
complicated so the inverse problem of determining junction parameters 
from the $I_{\rm max}(H)$ curve is very
difficult to solve for arrays or general current densities.
However, when the simple magnetic approximation of the 
$I_{max}(H)$ holds, it allows to extract information
on the sizes and positions of the junctions in an array
assuming $I_{max}(H)$ is a periodic and even function. This
is the purpose of this article. In particular we will
show how one can obtain a cosine profile and multi-cosine profile
from a combination of simple 3 junction arrays.
We will indicate what parameters can be obtained from a general
$I_{max}(H)$ profile. After presenting the general model in
section 2, we introduce the magnetic approximation and give its
properties in section 3. Section 4 discusses the inverse problem
for a three junction array. In section 5 we design the device 
from a general $I_{max}(H)$ and conclude in section 6.

\section{The model}

\begin{figure}
\centerline{\epsfig{file=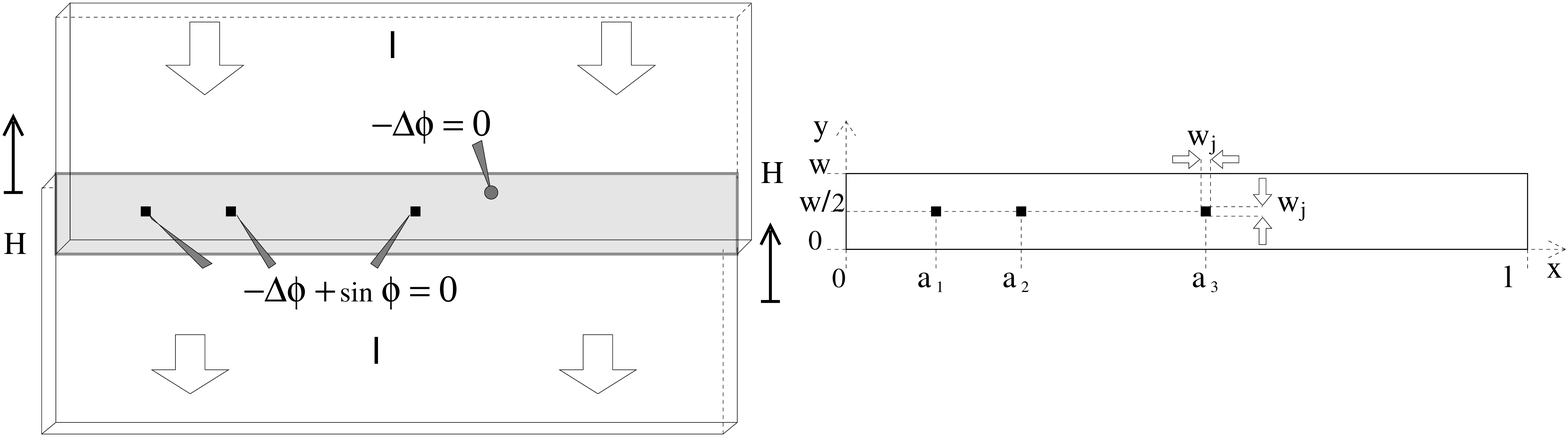,height=5cm,width=\linewidth,angle=0}}
\caption{The left panel shows the top view of a superconducting
microstrip line containing three Josephson junctions,
$H,I$ and $\phi$ are respectively the applied magnetic field, current and the
phase difference between the two superconducting layers. The phase difference
$\phi$ between the two superconducting layers satisfies
$-\Delta \phi=0$ in the linear part and
$-\Delta \phi+\sin(\phi)=0$ in the Josephson junctions. The right panel
shows the associated 2D domain of size $l \times w$ containing $n=3$ junctions
placed at the positions $y=w/2$ and $x=a_i,~i=1,n$.}
\label{f1}
\end{figure}

The device we model (see Fig.~\ref{f1}) is a so-called microstrip
cavity (grey area in Fig.~\ref{f1}) between two superconducting layers
containing small regions (junctions) where the oxide layer is
very thin ($\sim$ 10 Angstrom) enabling Josephson coupling 
between the top and bottom superconductors. The dimensions of 
the microstrip are about
100 $\mu$m length and 20 $\mu$m width and the length and width
of the junctions is about $w_j=1$ $\mu$m. In the static regime, the 
phase difference $\varphi$ between
the top and bottom superconducting layers obeys 
the following semilinear elliptic partial differential equation \cite{cfv95}
\begin{equation}\label{2dsg}
-\Delta\varphi+g(x;y)\sin\varphi = 0,
\end{equation}
where $g(x;y)=1$ in the Josephson junctions and 0 outside and where we
have neglected the difference in surface inductance between the junction and
passive region. 
This formulation guarantees the continuity of the normal gradient 
of $\varphi$, the
electrical current on the junction interface. The space unit is the
Josephson length $\lambda_J$, the ratio of the flux formed with the
critical current density and the surface inductance to the flux quantum
$\Phi_0$.

The boundary conditions representing an external current input $I$
or an applied magnetic field $H$ (along the y axis) are
\begin{equation} \label{bc} 
\left. \frac{\partial \varphi}{\partial y}\right|_{y=0}=-\frac{I }{2l}\nu,~ 
\left. \frac{\partial \varphi}{\partial y}\right|_{y=w}=\frac{I }{2l}\nu,~
\left. \frac{\partial \varphi}{\partial x}\right|_{x=0}=H-\frac{I}{2w}(1-\nu),~
\left. \frac{\partial \varphi}{\partial x}\right|_{x=l}=H+\frac{I}{2w}(1-\nu)
\end{equation}
where $0\leq \nu \leq 1$ gives the type of current feed. The case
$\nu = 1$ shown in Fig.~\ref{f1} where the current is only applied
to the long boundaries $y=0,w$ is called overlap feed while
$\nu=0$ corresponds to the inline feed.

We consider long and narrow strips containing a few small junctions
of area $w_i^2$ placed on the line $y=w/2$ and centered on $x=a_i, ~~i=1,n$
as shown in Fig.~\ref{f1}. Then we search $\varphi$ in the form
\begin{equation}\label{e2.2.1}
\varphi(x;y)=\frac{\nu I }{2L}\left(y-\frac{\omega}{2}\right)^2 +
\sum_{n=0}^{+\infty}\phi_n(x) \cos\left(\frac{n\pi y}{w}\right),
\end{equation}
where the first term takes care of the $y$ boundary condition.
For narrow strips $w<\pi$, only the first transverse
mode needs to be taken into
account \cite{cfgv96} because the curvature of $\varphi$ due
to current remains small. Inserting (\ref{e2.2.1}) into (\ref{2dsg})
and projecting on the zero mode we obtain the following
equation for $\phi_0$ where the index $0$ has been dropped for simplicity
\begin{equation}\label{e2.3}
-\phi^{\prime \prime} + g\left(x,{w\over 2}\right){w_i\over w} \sin \phi =
\nu \frac{\gamma}{l},
\end{equation}
and where $\gamma=I/w$ and the boundary conditions
$\phi^{\prime}(0)=H-(1-\nu)\gamma/2$,
and $\phi^{\prime}(l)=H+(1-\nu)\gamma/2$. The factor
$w_i/w$ is exactly the "rescaling" of $\lambda_J (=1)$ into
$\lambda_{eff} = \sqrt{w \over w_j}>1 $ due to the presence of the
lateral passive region \cite{cfv96}.

As the area of the junction is reduced, the total
Josephson current is reduced and tends to zero.
To describe small junctions where the phase variation
can be neglected but which can carry a significant current,
we introduce the following function $g_h$
\begin{equation}\label{gh}
g_h(x) = {w_i\over 2 h}, ~~~{\rm for}~ a_i-h<x<a_i+h,~~{\rm and}~~g_h(x) =0~~
~~{\rm elsewhere},\end{equation}
where $i=1,..n$.
In the limit $h\rightarrow 0$ we obtain our final delta function
model \cite{cl05}
\begin{equation}\label{e3.1}
-\phi^{\prime\prime}+\sum_{i=1}^n d_i \delta(x-a_i)\sin\phi = \nu {\gamma \over l} ,
\end{equation}
where $d_i = w_i^2/w$ and the boundary conditions
are $$ \phi^{\prime}(0) = H-(1-\nu)\gamma/2,   ~~
\phi^{\prime}(l) = H+(1-\nu)\gamma/2. $$
Despite its crude character the delta function approximation
is a good model for arrays with short junctions as long as the
magnetic field is small compared to $1/w_i$ where $w_i$ is the
size of the junctions \cite{bcls06}. It allows
simple calculations and in depth analysis that are 
out of reach for the 2D full model.
In addition when $d_i<0$ the model can describe so-called 
$\pi$-junctions. For these, the tunneling current
is $\sin(\phi+\pi)= -\sin(\phi)$ instead of the usual sine term
in the second Josephson equation (\ref{e1.1}).
This new type of coupling occurs in some materials \cite{Kirtley,Razianov} 
and it is hoped to be incorporated in the design of arrays.
It is then natural to associate negative $d_i$ coefficients to
$\pi$ junctions in the device.

We have the following properties.
\begin{enumerate}
\item Integrating twice (\ref{e3.1}) shows that the solution
$\phi$ is continuous at the junctions $x=a_i, ~i=1,\dots n$.
\item Almost everywhere (in the mathematical sense),
$-\phi^{\prime\prime}(x)= \nu \gamma/l$,
so that outside the junctions, $\phi$ is a piece-wise second degree
polynomial,
\begin{equation}\label{rem1}
\phi(x) = -\frac{\nu \gamma}{2 l}x^2 + B_i x + C_i~,~~ \forall x \in ]a_i;a_{i+1}[.
\end{equation}
\item At each junction ($x=a_i$), $\phi^{\prime}$ is not defined, but
choosing $\epsilon_1 > 0$, and $\epsilon_2 > 0$, we note that
$$\lim_{\epsilon_1 \rightarrow 0}\lim_{\epsilon_2 \rightarrow 0}
\int_{a_i-\epsilon_1}^{a_i+\epsilon_2} \phi^{\prime\prime}(x) dx =
\int_{a_i^-}^{a_i^+}\phi^{\prime\prime}(x)
dx= \left[\phi^{\prime}(x)\right]_{a_i^-}^{a_i^+}.$$
Since the phase is continuous at the junction $x=a_i$,
$\phi_i \equiv \phi(a_i)$ we get 
\begin{equation}\label{rem2}
\left[\phi^{\prime}(x)\right]_{a_i^-}^{a_i^+} = d_i\sin(\phi_i)\;.
\end{equation}
\item Integrating (\ref{e3.1}) over the whole domain,
$$\left[\phi^{\prime}\right]_0^l= \int_0^l \phi^{\prime\prime} dx =
\sum_{i=1}^n d_i \sin(\phi_i)-\nu \gamma~,$$ and taking into account the
boundary conditions, we obtain
\begin{equation}\label{rem3}
\gamma = \sum_{i=1}^n d_i\sin(\phi_i),
\end{equation}
which indicates the conservation of current.
\end{enumerate}
In \cite{cl06}, we developed two ways to find 
the $\gamma_{max}(H)$ curve
for the device using these properties, see the Appendix 
"Piece-wise polynomial" for more details on the solution of the problem. 
The most useful property in \cite{cl06} for this 
study is the magnetic approximation of the $\gamma_{max}(H)$ 
curve.

\section{The magnetic approximation}

Since $\left[\phi^{\prime}\right]_{a_i^-}^{a_i^+}=d_i \sin(\phi_i)$
(remark \ref{rem2}) and $\gamma \leq \sum_i d_i$, we notice that for small $d_i$,
$\phi$ tends to the linear function $\phi(x) = Hx + c$. Starting from
$\phi(x) \equiv Hx + c$, it is simple to find the $\gamma_{max}(H)$ curve.
This is what we call the magnetic approximation. We generalize here
what was done in \cite{mghg} for arrays of equidistant junctions.
We have shown in \cite{cl06} that 
the $\gamma_{max}(H)$ curve of (\ref{e3.1}) tends to it 
when $d_i$ tends to zero. In experiments the $d_i$ coefficients are small 
enough so that this approximation is justified and provides
a quantitative estimate of the $\gamma_{max}(H)$ curve \cite{bcls06}. For 
inhomogeneous arrays of many junctions, this curve is complex and
even in this case the approximation is very good.

Since $\phi(x) = Hx + c$ then $\gamma=\sum_i d_i \sin(Ha_i+c)$.
To find the $\gamma_{max}(H)$ curve of the magnetic approximation, we take the
derivative of
\begin{equation}\label{e3.3.1}
\gamma=\left(\sum_{i=1}^n d_i\sin(Ha_i)\right)\cos(c) + 
\left(\sum_{i=1}^n d_i \cos(Ha_i)\right) \sin(c) \equiv A \cos(c) + B \sin(c), 
\end{equation}
with respect to $c$ where we have isolated the factors $A,B$. The values of $c$ 
such that $\partial \gamma/\partial c = 0$ are
\begin{equation}\label{cmax}
c_{max}(H) = \arctan\left(\frac{\sum_{i=1}^n d_i \cos(Ha_i)}{\sum_{i=1}^n d_i
\sin(Ha_i)}\right)\;,
\end{equation}
and as we want a maximal (not only an extremal) current 
we obtain:
\begin{equation}\label{magn.app.}
\gamma_{max}(H) = \left|\sum_{i=1}^n d_i
\sin\left(Ha_i + c_{max}(H)\right)\right|\;.
\end{equation}
Now, we focus on the case where $c_{max}(H)$ is not defined. In this
case, considering previous equation (\ref{cmax}), we obtain
$\sum_{i=1}^n d_i \sin(Ha_i)=0=A$. From $A\sin(c)= B \cos(c)$ we obtain
$\cos(c)=0$ or $B=0$. Note that $\cos(c)=0$ imply $\gamma = 0$, in the
other hand, $B=0$, imply $\partial \gamma/\partial c = 0$ whatever
the value of $c$. So, $\gamma$ is constant, and $\gamma_{max}=A=0$.
\begin{figure}
\centerline{\epsfig{file=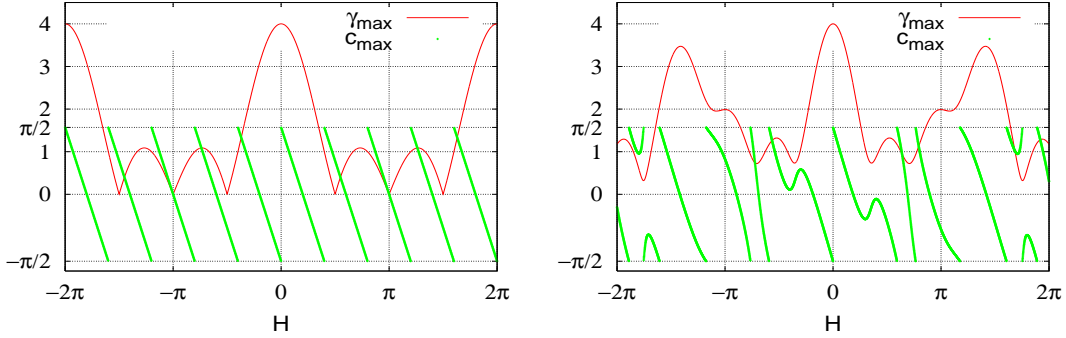,height=15 cm,width=5 cm,angle=270}}
\caption{We plot $\gamma_{max}(H)$ and $c_{max}(H)$ for two four
junction devices. In the left panel, we have a circuit with
equidistant junctions of equal $d_i=1$.
For the right panel, the junctions are such that $l_1 =\frac{5}{2},~ l_2 = \frac{5}{3},
~l_3 = 1$, with $d_1=1.2$, $d_2=1.5$, $d_3=0.5$, $d_4=0.8$. 
Notice that $c_{max}(H)$ varies in a complicated way for a non 
uniform device.}
\label{f2}
\end{figure}
We plot $\gamma_{max}(H)$ and $c_{max}(H)$ in Fig.(\ref{f2}), 
for a uniform device and for a non uniform one. In the second case, we have 
chosen the position of the junction to have a long 
period ($H_p=12\pi$).
We notice that the length $l$ of the device 
does not appear in eq. (\ref{magn.app.}). 

In order to study the function (\ref{magn.app.}), we start with a
few definitions.
\begin{description}
\item[\mbox{\boldmath $l_i$}:] 
We define the distance between consecutive junctions:
$$l_i=a_{i+1}-a_i~.$$
\item[Junction unit:] We call a junction unit the set of distances
between junctions. We denote it 
$$\{l_1;l_2;...;l_{n-1}\}~.$$ We define the position of the junction unit as $a_1$,
the position of the first junction.
\item[\mbox{\boldmath $l_b$}:]
For an $n$-junction device, $l_b=a_n-a_1$ is junction unit length.
\item[Symmetric unit:] We call a symmetric unit, an $n$-junction circuit 
such that
$$\forall i \in \{1;\dots;n\},~~
\frac{l_b}{2} - a_i = a_{n-i} - \frac{l_b}{2}\;,~~
{\rm and}~~ d_{i}=d_{n-i}\;.$$
\end{description}
In the Appendix we prove the following three Propositions
\begin{description}
\item [Invariance by translation] (\ref{thm1}): 
$\gamma_{max}(H)$ does not depend on the
position of the junction unit.
\item [Parity of the $\gamma_{max}$ curve] (\ref{thm2}):
$\gamma_{max}(H)$ is an even function of $H$.
\item [Solution for a symmetric device] (\ref{thm3}):
For a symmetric junction unit such that $a_n = -a_1$, $c_{max}(H)=\pm \pi/2$
(this is not obvious from fig.(\ref{f2})).
\end{description}
In these propositions, we establish the most important result 
of this article. The $\gamma_{max}$ curve for a symmetric junction unit can
be calculated simply by centering the junction unit so that $c_{max}(H)=\pm \pi/2$.
More precisely, consider an $n+1$ symmetric junction unit where $n$ is even.
We can always choose this by setting the $d$ of the central junction $(a_1+a_n)/2$ to 0.
Then we shift the junction unit by $a_{(n+1)/2}$ so that the central junction
is placed at $x=0$. We relabel the junctions by setting $i' = i -(n+1)/2$. Then
the central one is for
$i=0$, the first one to the right is $i=1$, 
the first one to the left is $i=-1$ \dots so that the equation (\ref{magn.app.})
becomes
\begin{equation}\label{gamma_sym}
\gamma_{max}(H) = \left|d_0+2\sum_{i=1}^{(n+1)/2} d_i\cos(H a_i)
\right|\;,
\end{equation}
where we omitted the primes. In the rest of the article we will
consider the array to be symmetric.

\section{The direct problem for $\gamma_{max}$}
\begin{figure}
\centerline{\epsfig{file=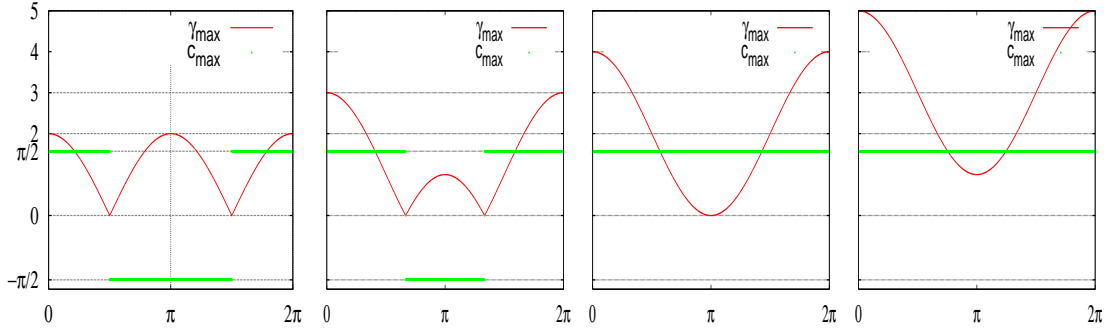,height=\linewidth,width=5 cm,angle=270}}
\caption{Curves $\gamma_{max}$ and $c_{max}$ versus $H$ for three
junction devices, from left to right $d_2=0$, $d_2=1$,
$d_2=2$ and $d_2=3$. For all panels, $a_1=-1$, $a_2=0$, $a_3=1$ and
$d_1=d_3=1$.} 
\label{f3}
\end{figure}

\subsection{A device such that $\gamma_{max}(H)=\cos(H)$}

In Fig.(\ref{f3}) we present from left to right the $\gamma_{max}(H)$
for a SQUID (2 junctions), a uniform 3 junction unit, a $d_1=1=d_3, ~d_2=2$ (termed 1-2-1) 
junction unit that
is discussed in \cite{Likharev} and a 1-3-1 junction unit. 
In all cases the junctions are equidistant.
The first two panels, represent well known devices.

Applying eq.(\ref{gamma_sym}) to the following case $d_2=2$, we 
obtain,
\begin{equation}\label{cos}
\gamma_{max}(H) = |d_0 + d_1 \cos(H a_1)| = 2 + 2 \cos(H a_1)\;.
\end{equation}
This is an exact cosine function shifted by a constant. 
With the last case $d_2=3$, we obtain 
$\gamma_{max}(H) = 3 + 2 \cos(H a_1)$.

Comparing all the panels we understand the role of 
the central junction. We can have an exact representation
of $\gamma_{max}$ for this type of circuit, if we imagine $\gamma_{max}$ 
as an absolute value of a simple $\cos$ function translated by
the value $d_0$ (which is equal to zero if there is no junction).
Eq.(\ref{gamma_sym}) shows that we can sum cosine functions, with a
chosen amplitude and period. Remark that if $d_0=-2$ ($\pi$ junction)
then $\gamma_{max}(H) = 2 -2 \cos(H a_1)$.

\subsection{A multi-cosine $\gamma_{max}(H)$}

\begin{figure}
\begin{minipage}{0.49\linewidth}
\epsfig{file=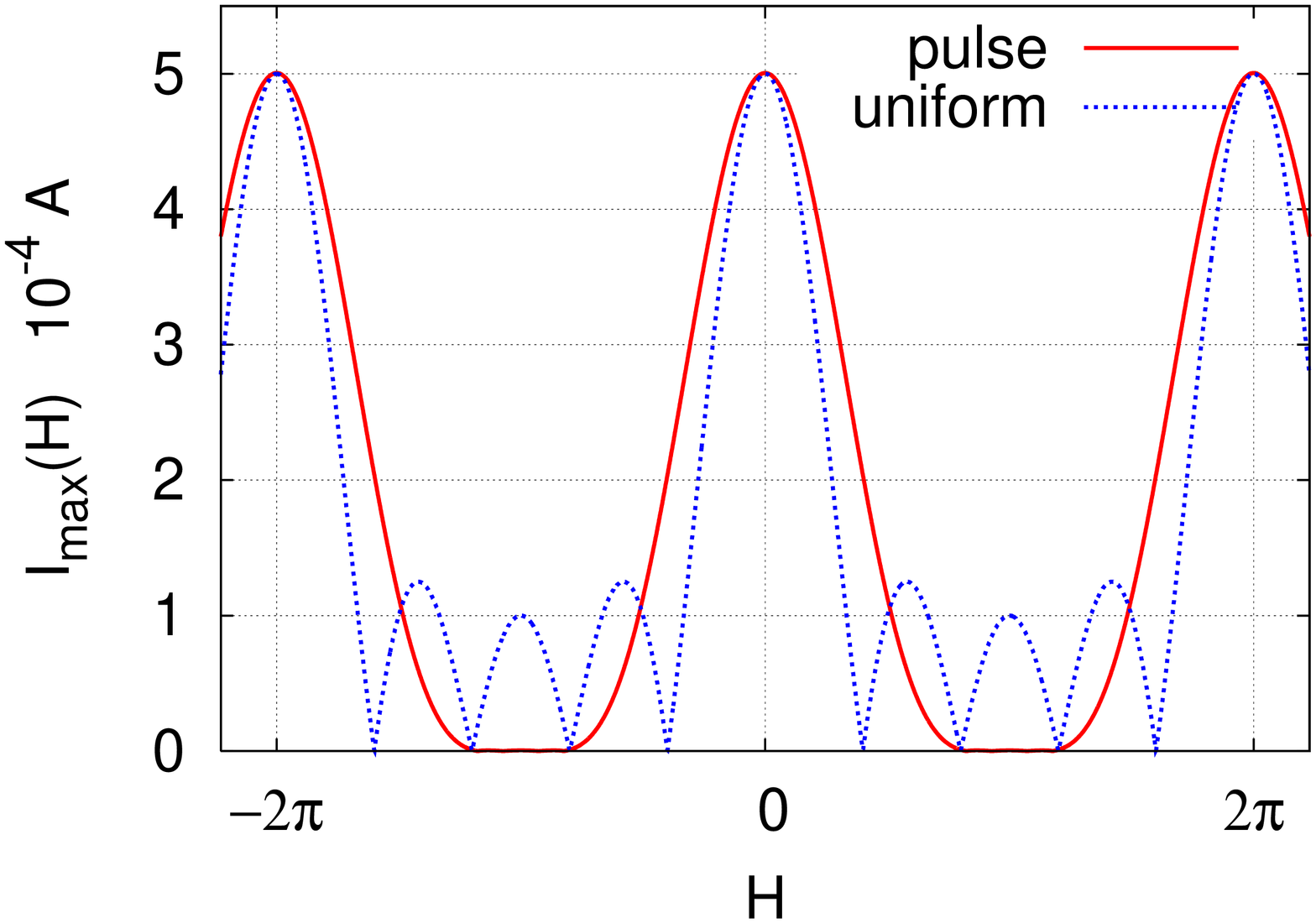,height=\linewidth,angle=270}
\end{minipage}
\begin{minipage}{0.49\linewidth}
\epsfig{file=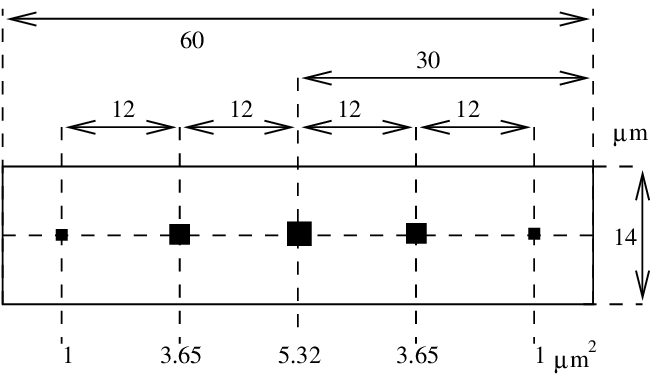,width=\linewidth,angle=0}
\end{minipage}
\caption{Left panel: plot of $\gamma_{max}(H)$ for 
a symmetric five equidistant 
junction device. The continuous line corresponds to different $d_i$ 
while the dashed line is for equal $d_i$. See the text for the parameter
values. The right panel shows the corresponding device.}
\label{f4}
\end{figure}

For arrays with more than two junctions, experimentalists can play on the set of distances $l_i$ separating the junctions as well as on the
strength $d_i$ (proportional to the area) of each junction.
We now show the influence of each set of parameters starting from the
$d_i$'s. Fig. \ref{f4} presents on the left panel
$\gamma_{max}(H)$ for a symmetric set of 5 equidistant
junctions $a_1=1,~a_2=2$. The dashed line corresponds 
to $d_i=1,~i=-2 \dots 2$ giving a maximum current 
of 5. Here one sees the typical interference pattern 
between the main bumps. The small oscillations can be eliminated
by choosing $d_0=1.82025, ~d_1= d_{-1}=1.25$ and 
$d_2= d_{-2}=0.3425$ as seen from the continuous line on the
left panel of Fig. \ref{f4}. This "pulse" profile could be very useful
for specific applications because of the large region where
$\gamma_{max}(H)=0$. The right panel of Fig. \ref{f4} presents what 
would be the device for this set of $a_i$ and $d_i$. We 
chose a critical current density
of $10^4 A~ cm^{-1}$ so that $\lambda_J \approx 10 \mu m$.
We chose a transverse width $w=14\mu m$. Assuming the area of 
the smallest junction to be $1 \mu m^2$, we get the scheme
shown, where the central junction has an area 5.32 $\mu m^2$.

\begin{figure}
\begin{minipage}{0.49\linewidth}
\epsfig{file=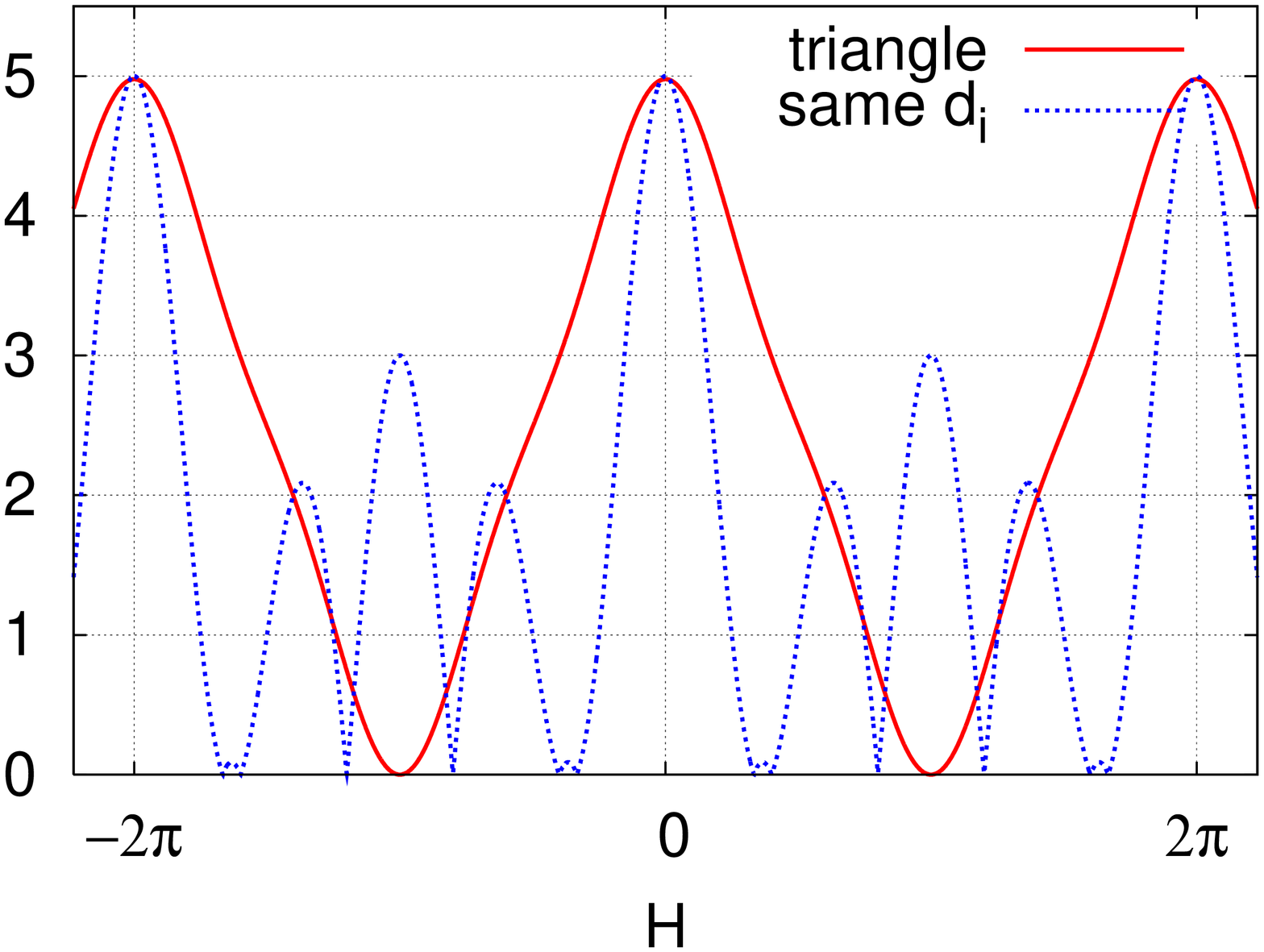,height=\linewidth,angle=270}
\end{minipage}
\begin{minipage}{0.49\linewidth}
\epsfig{file=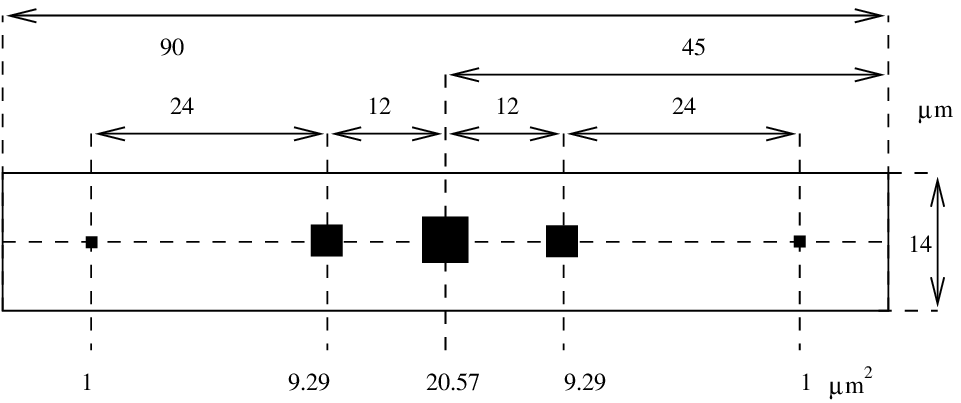,width=\linewidth,angle=0}
\end{minipage}
\caption{Left panel: plot of $\gamma_{max}(H)$ for a symmetric five junction
device where the junctions are not equidistant, $a_1=1$ and $a_2=3$. 
The continuous line corresponds to different $d_i$ while the dashed line
is for equal $d_i$. The right panel shows the corresponding device.}
\label{f5}
\end{figure}

The second parameter that can be changed is the position $a_i$ of
each junction in the array. As an example in Fig. \ref{f5}
we show in the left panel the so-called "triangle" $\gamma_{max}(H)$ obtained
by setting $a_1=1,~a_2=3$, $d_0=2.4888, ~d_1=1.1234$ and $d_2=0.121$.
The dashed line presents $\gamma_{max}(H)$ for equal strengths. 
Changing the $d_i$'s allows to eliminate the oscillations in the 
minima of $\gamma_{max}(H)$ and obtain an almost linear behavior. The 
right panel shows the 
arrangement of the junctions in the microstrip. We have chosen the 
same physical parameters as for Fig. \ref{f4}.
Now we can design all devices which have a $\gamma_{max}(H)$ curve as
a sum of $d_i \cos(a_i x)$, with $d_i$ positive. We can notice that if
$a_i/a_1 \in \mathbb{R\backslash Q}$, we can construct a non
periodic $\gamma_{max}(H)$. 

\section{The inverse problem for a given $\gamma_{max}(H)=\gamma_g(H)$}

We now show how to design an $n+1$ junctions
circuit ($n$ is an even integer) to obtain a given $\gamma_g(H)$ curve. 
The formula (\ref{gamma_sym}) can be used to solve this inverse problem
using cosine Fourier transforms. To avoid ambiguities we assume a
symmetric array and a positive and periodic $\gamma(H)$.

We have the following result.
\begin{proposition}[Solution of inverse problem for $\gamma_{max}(H)$]
\label{thm4}
Assume a $\gamma_g(H)$ even, periodic of period $H_p$ and strictly positive.
The array is harmonic and the positions of the junctions 
are given by $a_i = i 2 \pi /H_p $ where $i$ is an integer. 
Their strengths $d_i$ are given by
\begin{equation}\label{coef_four}
d_{-i}=d_i=\frac{1}{H_p}\int_0^{H_p} \gamma_g(H)\cos(H a_i) dH~,~~
\forall i \in \{0,\dots,n/2\}~.
\end{equation}
\end{proposition}
This gives the positions $a_i$ and coefficients $d_i$ of an array that
will have a $\gamma_{max}(H)$ that is the truncation to order $n$ of
the cosine Fourier series of $\gamma_g(H)$.

To gain insight into the problem let us first review the "pulse"
example studied in the previous section. Assume 
$\gamma_g(H)$ to be the $2\pi$ periodic extension of $e^{-\alpha H^2}$ where 
$\alpha $ is large enough.  The coefficients $d_i$ are given by
\be\label{dkpulse}
d_i = 
{1 \over 2\pi} \int_0^{2\pi} e^{-\alpha H^2} \cos (H i)~dH 
+ {1 \over 2\pi} \int_0^{2\pi} e^{-\alpha (2\pi-H)^2} \cos (H i)~dH  
={1 \over \pi} \int_0^{2\pi} e^{-\alpha H^2} \cos (H i)~dH .
\ee
These Fourier coefficients decay exponentially as expected \cite{carslaw}
because $\gamma_g(H)$ is $C^\infty$ over the interval $[0, 2\pi]$ and
satisfies the boundary conditions. This means that $i\le 2$ is enough to
get a good approximation of $\gamma_g(H)$. In fact Fig. \ref{f4} corresponds
to $\gamma_g(H)\approx 5 e^{-0.6 H^2}$ and the
formula (\ref{dkpulse}) gives the values 
$d_0=1.82025, ~d_1= d_{-1}=1.25$ and 
$d_2= d_{-2}=0.3425$ that were obtained in the previous section. The
next coefficients $d_3=0.043$, $d_4=0.0023$ are very small
and can be neglected.

Let us now consider a square $\gamma_{max}(H)$ curve which
could make a very fine magnetic detector because of its
strong response over a given interval and zero response
elsewhere. For that we assume the the square profile
\begin{equation}\label{creneau}
\gamma_g(H)=1 ~~ {\rm for} ~~\pi\left(1-{h_1 \over 2}\right) < H< \pi\left(1-{h_1 \over 2}\right)
~~{\rm and}~~  0 ~~{\rm elsewhere},
\end{equation}
and extend it periodically every $2 \pi = H_p$.
To compute the parameters $a_i$ and $d_i$, we apply the previous result
(see eq.(\ref{coef_four})) to obtain
\begin{equation}
d_i={1 \over 2 \pi} \int_{0}^{2\pi} \gamma_g(H) \cos\left({i \pi H \over 2 \pi}\right) dH
= {2 \over i \pi }\sin \left (i \pi {h_1 + h_2\over 2}\right) 
\cos \left({i \pi \over 2 } + {h_2-h_1 \over 2 }\right) 
\end{equation}
This gives the following values of $d_i$ for $h_1=h_2=h$
$$\begin{array}{|c|c|c|c|c|c|c|}\hline
i   & 0 & 1 & 2 & 3 & 4 & \dots \\ \hline
a_i & 0 & 1 & 2 & 3 & 4 & \dots \\ \hline
2 \pi d_i & 2 h & 0 & -{\sin (2 \pi h) \over \pi}  & 0 & {\sin (4 \pi h) \over 2 \pi} &  \dots \\ \hline 
\end{array}$$
\begin{figure}
\centerline{\epsfig{file=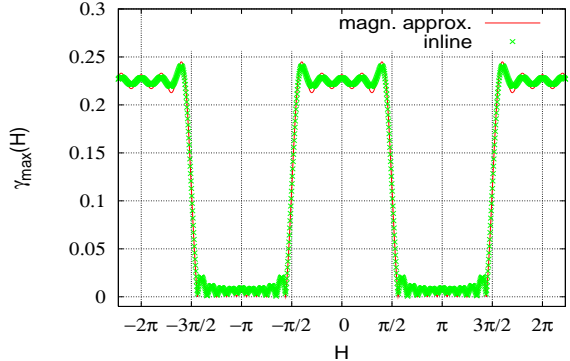,height=8 cm,width=5 cm,angle=270}}
\caption{We compare the $\gamma_{max}(H)$ of the magnetic approximation 
to the exact solution of equation (\ref{e3.1}) for $\nu=0$ (inline
current feed. The parameters are given in the text.}
\label{f6}
\end{figure}
Note the decay in $1 /i$ of $d_i$ because $\gamma_g(H)$ is only
continuous. Another interesting fact is that some $d_{i}$ are
negative so that some junctions are $\pi$ junctions.
So we obtain an array of eleven junctions whose positions are
given above together with their strengths $d_i$ (positive for
a normal junction and negative for a $\pi$ junction).

In Fig. \ref{f6} we plot the magnetic approximation 
and the exact solution of equation (\ref{e3.1}) for $\nu=0$ 
(this case is called inline current feed, see the section 
"Piece-wise polynomial"
in the Appendix and for more details \cite{cl06}).
The values are $l=20$, the junction unit is shifted by $10$, the 
Josephson characteristic length is $\lambda_J=5.6 \mu m$ so that 
all $d_i$ are multiplied by $0.035714285$.  

We see that for this type of junction (about $1 \mu m^2$ of area) 
inline current feed for (\ref{e3.1}) and magnetic approximation give
close results. 
Differences appear when the maximal current is larger
but the Gibbs phenomena is less important in the solution 
of the equation (\ref{e3.1}) than magnetic approximation.

\section{Conclusion}

Using a simple approximation, we introduced a method to
design a symmetric array of Josephson junctions which has
a specific $\gamma_{max}(H)$ static response. The sizes
of the junctions are given by the coefficients $d_i$ of the cosine
Fourier transform of $\gamma_{max}(H)$.  Their position is
$a_i= i 2 \pi/H_p$ where $H_p$
is the period of $\gamma_{max}(H)$.
We use $2n+1$ junctions to obtain a curve formed with
$n$ Fourier coefficients. 

This work follows closely the article \cite{cl06}, where all
the mathematical results were established, in particular the convergence
of the solutions of the full problem (\ref{e3.1}) to the ones
obtained in the magnetic approximation. 
There we show that the overlap current feed can cause a non even 
$\gamma_{max}(H)$ (see proposition "Magnetic shift" in \cite{cl06}). 

If we are in the region of validity of our original model, ie
the magnetic field is small and the distance between junctions is
larger than their size, then we can design a device for
a given static response.

{\bf Acknowledgements}

The authors thank Faouzi Boussaha and Morvan Salez for useful
discussions.
L. L. thanks Delphine and Damien Belmessieri for their comments.

\section{Appendix}
\subsection{Propositions}

\begin{proposition}[Invariance by translation]
\label{thm1} The $\gamma_{max}(H)$ curve obtained within the
magnetic approximation
does not depend on the position of the junction unit. \end{proposition}

\begin{proof} Let us assume two devices with the same 
junction unit $\{l_1;l_2;...;l_n\}$ with the first junction
placed respectively at $x=a_1$ and $x=a_1+c$.
We note $\gamma^1_{max}(H)$
(respectively $\gamma^2_{max}(H)$) the $\gamma_{max}(H)$ curve of the
first (respectively the second) device. In the same way we note
$c_{max}^1(H)$ (respectively $c_{max}^2(H)$) the $c_{max}(H)$
function of the first (respectively the second) device.
$$\gamma^2_{max}(H) = \left|\sum_{i=1}^n d_i
\sin\left(Ha_i+ Hc+c_{max}^2(H)\right)\right|\;,$$
we note $c^1(H)=Hc+c_{max}^2(H)$. As we do not know if $c^1(H)=
c_{max}^1(H)$, $c_{max}^1(H)$ being the best value (if it exists)
to obtain the maximal $\gamma$. Consequently,
$\gamma^2_{max}(H)\leq \gamma^1_{max}(H)$.

On the other side, considering
$$\gamma^1_{max}(H) = \left|\sum_{i=1}^n d_i
\sin\left(H(a_i+c)+c_{max}^1(H)-Hc\right)\right|\;,$$
noting $c^2(H)=c_{max}^1(H)-Hc$ and using the previous argument, we
obtain $\gamma^2_{max}(H)\geq \gamma^1_{max}(H)$.
From the two previous inequalities, we obtain:
$\gamma^1_{max}(H) = \gamma^2_{max}(H)$.
\end{proof}

Thus, the $\gamma_{max}(H)$ curve for the magnetic approximation depends only
on the junction unit. 

\begin{proposition}[Parity of the $\gamma_{max}$ curve]\label{thm2}
For all devices, $\gamma_{max}(H)=\gamma_{max}(-H)$.
\end{proposition}
\begin{proof}
Since $\sin$ and $\arctan$ are odd functions and $\cos$ is an even function
then $c_{max}(-H)=-c_{max}(H)$ (see (\ref{cmax})).
Finally,
$$\gamma_{max}(-H)=\left|\sum_{i=1}^n d_i \sin\left(-Ha_i - c_{max}(H)
\right)\right|=\gamma_{max}(H)\;.$$\end{proof}

Notice that $c_{max}$ is an odd function and
$\gamma_{max}$ is an even function (see Fig.(\ref{f2}) ).

\begin{proposition}[Particular solution for symmetric device] \label{thm3}
For all symmetric units such that $a_n = -a_1$, $c_{max}(H)=\pm \pi/2$.
\end{proposition}
\begin{proof} To see this, relabel the junctions so that the central one 
corresponds to $i=0$, the 1st on the left to $i=-1$, the 1st on the right to 
$i=+1$.. Using the first proposition we can shift the junction unit so that
$a_0=0$. Then the total current is
\begin{eqnarray*}
\gamma(H) &=&\sum_{i=-n/2}^{n/2} d_i\sin(Ha_i+c) \\
&=& d_0 \sin(c) + 2 \sum_{i=1}^{n/2} d_i\sin(Ha_i+c) \\
&=& \sin(c) \left( d_0 + 2 \sum_{i=1}^{n/2} d_i \cos(Ha_i) \right ) 
\end{eqnarray*}
Then $c_{max}=\pi/2$ when $d_0 + 2 \sum_{i=1}^{n/2} d_i \cos(Ha_i)
\geq 0$ and $c_{max}=-\pi/2$ otherwise. Thus, 
\begin{equation}\label{sol1} 
\gamma_{max}(H) = \left|d_0+2\sum_{i=1}^{(n+1)/2} d_i\cos(H a_i)
\right|\;.
\end{equation}
\end{proof}

\subsection{Piece-wise polynomial }

Let $\phi$ be a solution of (\ref{e3.1}) and $\phi_1=\phi(a_1)$.
From remark (\ref{rem1}), $\phi$ is a polynomial by parts. We define
$P_{i+1}(x)$ the second degree polynomial such that $P_{i+1}(x)=\phi(x)
~~{\rm for}~~ a_{i} \le x \le a_{i+1}$.
Using the left boundary condition we can specify $\phi$ on $[0;a_1]$:
\begin{equation}\label{rec1}
P_1(x)=-\frac{\nu \gamma}{2 l}\left(x^2-a_1^2\right)
+ \left(H-\frac{1-\nu}{2}\gamma\right)(x-a_1) + \phi_1
\end{equation}
At the junctions (\ref{rem2}) tells us that $\forall k \in \{1;\dots;n\}$,
\begin{equation}\label{jump_a}
P_{k+1}^{\prime}(a_k) - P_k^{\prime}(a_k)= d_k \sin(P_k(a_k)).
\end{equation}
Considering that $\phi^{\prime \prime}=-\nu \gamma / l$ on each interval,
the previous relation and the continuity
of the phase at the junction, we can give a first expression for $P_{k+1}$,
\begin{equation}\label{recn}
P_{k+1}(x)=-\frac{\nu \gamma}{2l}(x-a_k)^2  + \left[P^{\prime}_k(a_k)+d_k \sin
P_k(a_k) \right](x-a_k)+P_k(a_k).
\end{equation}
So, $\phi$ is entirely determined by $\phi_1$, $\gamma$ and $H$.

The polynomials (\ref{rec1}) and (\ref{recn}) establish existence and shape
of $\phi$ at junctions. Let see boundary conditions. The first,
$$\phi^{\prime}(0)=P_1^{\prime}(0)=H-(1-\nu)\gamma/2$$
is true by construction; the second (for $n$ junction circuit) is:
\begin{equation}\label{rbc}
P^{\prime}_{n+1}(l)=H +(1-\nu)\frac{\gamma}{2},
\end{equation}
is true only for solutions of Eq.(\ref{e3.1}). At $H$ given,
solutions of Eq.(\ref{rbc}) define a relation between $\phi_1$ 
and $\gamma$.

So, the maximal current solution depend on $\phi_1$ and $\gamma$, and 
Eq.(\ref{rbc}) is the constraint for search of $\gamma_{max}(H)$. 
As solutions $\phi$ are defined at $2\pi$ almost (see equation (\ref{e3.1})), 
we can assume $\phi_1\in[-\pi;\pi]$. In other hand, Rem.(\ref{rem3}) teach us 
that $\gamma_{max}\in[0;\sum_i d_i]$, so we take 
$\gamma \in \left[0;\sum_i d_i\right]$. To solve this problem with Maple \copyright, 
we plot implicit function (the constraint) of two variables $\phi_1$ and 
$\gamma$, with $H$ and $\nu$ fixed. Let us note all variables:
\begin{equation}\label{impli}
\left .P^{\prime}_{n+1}\right|_{x=l}
(\phi_1,\gamma,\nu,H)-H-\frac{1-\nu}{2}\gamma=0.
\end{equation}
with $(\phi_1,\gamma)\in[-\pi;\pi]\times\left [ 0;\sum_i d_i \right ]$.
Lastly the program search in exhaustive way, the biggest value of
$\gamma$ of this implicit curve. Incrementing $H$, we obtain $\gamma_{max}(H)$.

This method has the advantage to converge to the global maximal $\gamma_{max}$ \cite{cl06}.

\end{document}